\documentclass[apj]{emulateapj}

\def\ks{km~s$^{-1}$}
\def\ms{m~s$^{-1}$}
\def\msy{m~s$^{-1}$~yr$^{-1}$}
\def\mjup{M$_{\rm Jup}$}

\def\msun{M$_{\odot}$}
\def\rsun{R$_{\odot}$}
\def\lsun{R$_{\odot}$}
\def\msini{$M_P\sin i~$}
\def\chisq{$\sqrt{\chi^2_\nu}$}
\def\chis{$\chi^2_\nu$}

\def\feh{[Fe/H]}
\def\rphk{$\log R^\prime_{\rm HK}$}

\def\logg{$\log{g}$}
\def\teff{T$_{\rm eff}$}
\def\vsini{$V_{\rm rot}\sin{i}$}
\def\caii{Ca\,II}

\def\starB{HD\,4313}
\def\pB{356.0}
\def\peB{2.6}
\def\tpB{2454804}
\def\tpeB{80}
\def\eB{0.041}
\def\eeB{0.037}

\def\omB{76}
\def\omeB{80}
\def\kB{46.9}
\def\keB{1.9}
\def\msiniB{2.3}
\def\msinieB{0.2}
\def\arelB{1.19}
\def\areleB{0.03}
\def\rmsB{3.7}
\def\chisB{0.79}
\def\nobsB{29}
\def\mstarB{1.72}
\def\bvB{0.96}

\def\vmagB{7.83}

\def\mvB{2.2}
\def\mveB{0.3}
\def\vsiniB{2.76}
\def\ageB{2.0}
\def\ageeB{0.5}
\def\rstarB{4.9}
\def\rstareB{0.1}
\def\lstarB{14.1}
\def\lstareB{0.5}
\def\teffB{5035}
\def\loggB{3.4}
\def\feB{+0.14}
\def\dB{137}
\def\deB{14}
\def\starC{HD\,95089}
\def\pC{507}
\def\peC{16}
\def\tpC{2454983}
\def\tpeC{90}
\def\eC{0.157}
\def\eeC{0.086}

\def\omC{317}
\def\omeC{60}
\def\kC{23.5}
\def\keC{1.9}
\def\trendC{30.9}
\def\trendeC{1.9}

\def\msiniC{1.2}
\def\msinieC{0.1}
\def\arelC{1.51}
\def\areleC{0.05}
\def\rmsC{5.7}
\def\chisC{1.32}
\def\nobsC{22}
\def\mstarC{1.58}
\def\bvC{0.94}

\def\vmagC{7.92}

\def\mvC{2.1}
\def\mveC{0.3}
\def\vsiniC{2.74}
\def\ageC{2.5}
\def\ageeC{0.9}
\def\rstarC{4.9}
\def\rstareC{0.1}
\def\lstarC{13.5}
\def\lstareC{0.5}
\def\teffC{5002}
\def\loggC{3.4}
\def\feC{+0.05}
\def\dC{139}
\def\deC{16}
\def\starD{HD\,181342}
\def\pD{663}
\def\peD{29}
\def\tpD{2454881}
\def\tpeD{50}
\def\eD{0.177}
\def\eeD{0.057}

\def\omD{281}
\def\omeD{30}
\def\kD{52.3}
\def\keD{3.7}
\def\msiniD{3.3}
\def\msinieD{0.2}
\def\arelD{1.78}
\def\areleD{0.07}
\def\rmsD{4.7}
\def\chisD{1.14}
\def\nobsD{17}
\def\mstarD{1.84}
\def\bvD{1.02}

\def\vmagD{7.55}

\def\mvD{2.2}
\def\mveD{0.2}
\def\vsiniD{3.04}
\def\ageD{1.8}
\def\ageeD{0.4}
\def\rstarD{4.6}
\def\rstareD{0.1}
\def\lstarD{12.0}
\def\lstareD{0.5}
\def\teffD{5014}
\def\loggD{3.4}
\def\feD{+0.26}
\def\dD{110.6}
\def\deD{7.5}
\def\starE{HD\,206610}
\def\pE{610}
\def\peE{13}
\def\tpE{2454677}
\def\tpeE{20}
\def\eE{0.229}
\def\eeE{0.058}

\def\omE{296}
\def\omeE{10}
\def\kE{40.7}
\def\keE{1.9}
\def\msiniE{2.2}
\def\msinieE{0.1}
\def\arelE{1.68}
\def\areleE{0.05}
\def\rmsE{4.8}
\def\chisE{1.06}
\def\nobsE{24}
\def\mstarE{1.56}
\def\bvE{1.01}

\def\vmagE{8.34}

\def\mvE{2.2}
\def\mveE{0.4}
\def\vsiniE{2.57}
\def\ageE{3}
\def\ageeE{1}
\def\rstarE{6.1}
\def\rstareE{0.1}
\def\lstarE{18.9}
\def\lstareE{0.6}
\def\teffE{4874}
\def\loggE{3.3}
\def\feE{+0.14}
\def\dE{194}
\def\deE{36}
\def\starF{HD\,136418}
\def\pF{464.3}
\def\peF{3.2}
\def\tpF{2455228}
\def\tpeF{10}
\def\eF{0.255}
\def\eeF{0.041}

\def\omF{12}
\def\omeF{10}
\def\kF{44.7}
\def\keF{1.9}
\def\trendF{-5.3}
\def\trendeF{1.7}

\def\msiniF{2.0}
\def\msinieF{0.1}
\def\arelF{1.32}
\def\areleF{0.03}
\def\rmsF{5.0}
\def\chisF{1.15}
\def\nobsF{25}
\def\mstarF{1.33}
\def\bvF{0.93}

\def\vmagF{7.88}

\def\mvF{2.7}
\def\mveF{0.2}
\def\vsiniF{0.17}
\def\ageF{4}
\def\ageeF{1}
\def\rstarF{3.4}
\def\rstareF{0.1}
\def\lstarF{6.8}
\def\lstareF{0.5}
\def\teffF{5071}
\def\loggF{3.6}
\def\feF{-0.07}
\def\dF{98.2}
\def\deF{5.6}
\def\starG{HD\,212771}
\def\pG{373.3}
\def\peG{3.4}
\def\tpG{2454947}
\def\tpeG{40}
\def\eG{0.111}
\def\eeG{0.060}

\def\omG{55}
\def\omeG{40}
\def\kG{58.2}
\def\keG{7.8}
\def\msiniG{2.3}
\def\msinieG{0.4}
\def\arelG{1.22}
\def\areleG{0.03}
\def\rmsG{5.8}
\def\chisG{1.29}
\def\nobsG{20}
\def\mstarG{1.15}
\def\bvG{0.88}

\def\vmagG{7.60}

\def\mvG{2.2}
\def\mveG{0.3}
\def\vsiniG{2.63}
\def\ageG{6}
\def\ageeG{2}
\def\rstarG{5.0}
\def\rstareG{0.1}
\def\lstarG{15.4}
\def\lstareG{0.5}
\def\teffG{5121}
\def\loggG{3.5}
\def\feG{-0.21}
\def\dG{131}
\def\deG{14}
\def\starH{HD\,180902}
\def\pH{479}
\def\peH{13}
\def\tpH{2454690}
\def\tpeH{130}
\def\eH{0.09}
\def\eeH{0.11}

\def\omH{300}
\def\omeH{100}
\def\kH{30.7}
\def\keH{3.7}
\def\trendH{135.4}
\def\trendeH{3.5}

\def\msiniH{1.6}
\def\msinieH{0.2}
\def\arelH{1.39}
\def\areleH{0.04}
\def\rmsH{3.3}
\def\chisH{0.98}
\def\nobsH{12}
\def\mstarH{1.52}
\def\bvH{0.94}

\def\vmagH{7.78}

\def\mvH{2.5}
\def\mveH{0.3}
\def\vsiniH{2.88}
\def\ageH{2.8}
\def\ageeH{0.7}
\def\rstarH{4.1}
\def\rstareH{0.1}
\def\lstarH{9.4}
\def\lstareH{0.5}
\def\teffH{5030}
\def\loggH{3.5}
\def\feH{0.04}
\def\dH{110}
\def\deH{10}

\def\sB{0.12}
\def\rphkB{-5.27}
\def\sC{0.13}
\def\rphkC{-5.22}
\def\sD{0.12}
\def\rphkD{-5.31}
\def\sE{0.14}
\def\rphkE{-5.23}
\def\sH{0.15}
\def\rphkH{-5.14}

\def\sF{0.14}
\def\rphkF{-5.19}
\def\sG{0.16}
\def\rphkG{-5.09}

\def\nplup{SEVEN}
\def\npllet{seven}

\begin{document}
\title{Retired A Stars and Their Companions IV. \\ \nplup\ Jovian Exoplanets
  from Keck Observatory$^1$}  

\author{John Asher Johnson\altaffilmark{2,3},
Andrew W. Howard\altaffilmark{4},
Brendan P. Bowler\altaffilmark{3},
Gregory W. Henry\altaffilmark{5},
Geoffrey W. Marcy\altaffilmark{4},
Jason T. Wright\altaffilmark{6},
Debra A. Fischer\altaffilmark{7},
Howard Isaacson\altaffilmark{3}
}

\email{johnjohn@astro.berkeley.edu}

\altaffiltext{1}{ Based on observations obtained at the
W.M. Keck Observatory, which is operated jointly by the
University of California and the California Institute of
Technology. Keck time has been granted by both NASA and
the University of California.}
\altaffiltext{2}{Department of Astrophysics,
  California Institute of Technology, MC 249-17, Pasadena, CA 91125}
\altaffiltext{3}{Institute for Astronomy, University of Hawai'i, 2680
  Woodlawn Drive, Honolulu, HI 96822} 
\altaffiltext{4}{Department of Astronomy, University of California,
Mail Code 3411, Berkeley, CA 94720}
\altaffiltext{5}{Center of Excellence in Information Systems, Tennessee
  State University, 3500 John A. Merritt Blvd., Box 9501, Nashville, TN 37209}
\altaffiltext{6}{Department of Astronomy \& Astrophysics, The
  Pennsylvania State University, University Park,  PA 16802}
\altaffiltext{7}{Department of Astronomy, Yale University, New Haven,  CT 06511}

\begin{abstract}
We report precise Doppler measurements of \npllet\
subgiants from Keck Observatory. All
\npllet\ stars show variability in
their radial velocities consistent with planet--mass companions in
Keplerian orbits. The host stars have masses ranging from $1.1 \leq
M_\star/M_\odot \leq 1.9$, radii $\rstarF \leq
R_\star/R_\odot \leq \rstarE$, and metallicities $-0.21
\leq$~\feh~$ \leq +0.26$. The planets are all more massive than
Jupiter (\msini~$ > 1$~\mjup) and have semimajor axes $a > 1$~AU. We
present millimagnitude  
photometry from the T3 0.4~m APT at Fairborn observatory for five of
the targets. Our monitoring shows
these stars to be photometrically stable, further strengthening the
interpretation 
of the observed radial velocity variability. The orbital characteristics
of the planets thus far discovered around former A-type stars are very
different from the properties of 
planets around dwarf stars of spectral type F, G and K, and suggests
that the formation and 
migration of planets is a sensitive function of stellar mass. Three of
the planetary systems show evidence of long-term, 
linear trends indicative of additional distant companions. These
trends, together with the high planet masses and 
increased occurrence rate, indicate that A-type stars are very
promising targets for direct imaging surveys.
\end{abstract}


\section{Introduction}

The field of exoplanetary science recently reached a major milestone
with the first direct-imaging detections of planetary systems
around main sequence stars\footnote{The planet candidate around the
  A-type star $\beta$~Pic announced by \citet{lagrange09} has not yet
  been confirmed by proper motion
  \citep{fitz09,lagrange09b}. Similarly, there
    exists no proper motion follow-up of the faint object imaged
    around 1RXS J160929.1-210524
    \citep{lafreniere08}.}. \citet{kalas08} detected a single  
planet-sized object with a semimajor axis $a \approx 120$~AU, orbiting
just inside of the dust belt around the nearby, young A4V star
Fomalhaut. The young A5V dwarf star HR\,8799 is orbited by a system of
three substellar objects with semimajor axes $a = \{24,38,68\}$~AU
\citep{marois08}. These remarkable systems share a number of
characteristics in common. Both host-stars are A-type dwarfs with
stellar masses $> 1.5$~\msun\ surrounded by debris disks, the planets
are super-Jupiters with masses $\lesssim 3$~\mjup, and the
companions orbit far from their central stars with unexpectedly large
semimajor axes (ranging from $20-120$~AU).  

That both systems were discovered orbiting A stars might at
first seem unlikely, given that A stars make up less than 3\% of the
stellar population in the Solar neighborhood and because the
star-planet contrast ratios are unfavorable compared to systems with
fainter, less massive central stars. However, in light of recent
discoveries from Doppler-based planet searches of massive stars 
it is becoming apparent that A dwarfs
may in fact be ideal target stars for direct
imaging surveys \citep{hatzes03, setiawan05, reffert06, sato07,
  nied07, liu08, dollinger09}. Measurements of the frequency of giant
planets around the ``retired''
counterparts of A-type dwarfs (subgiants and 
giants) have found that the occurrence of Jovian planets scales with
stellar mass: A-type stars ($M_\star \gtrsim 1.5$~\msun) are at least 5
times more likely than M dwarfs to harbor a giant planet 
\citep{johnson07b,bowler10,johnson10a}. And just like the current sample of
imaged planets,  Doppler-detected planets around retired A stars are more
massive \citep[][]{lovis07} and orbit farther from their stars than do planets
found around Sun-like, F, G and K (FGK) dwarfs \citep{johnson07, sato08b}. 

Indeed, there is strong evidence that the orbital characteristics of
planets around A stars are drawn from a statistical parent
population that is distinct from those of planets around FGK
dwarfs. \citet{bowler10} performed a statistical analysis of planets
detected in  
the Lick Subgiants Survey, which comprises 31 massive stars ($M_\star
\gtrsim 1.5$~\msun) monitored for the past 5 years . The mass-period
distribution of exoplanets around FGK dwarfs is typically
described by a double-power-law relationship, with the frequency of
planets rising toward lower masses and remaining flat in
logarithmic semimajor-axis 
bins from $\sim 0.05$~AU to $\sim5$~AU \citep{tab02,lineweaver03,
  cumming08, johnson09rev}. Based on the 7 planet detections from the
Lick survey, 
Bowler et al. concluded that the power-law indices of the distribution
of planets around A stars and Sun-like stars differ at the 4-$\sigma$
level; the planets in their sample all have \msini~$ > 1.5$~\mjup\ and
none orbit within 1~AU. However, their small sample size precluded a
determination of the exact shape of the mass-period
distribution. Fortunately, given the $26^{+9}_{-8}$\% occurrence rate
measured from the Lick survey, it will not take long to build a
statistical ensemble comparable to the collection of planets around
less massive stars.  

To increase the collection of planets detected around massive stars,
and to study the relationships among the characteristics of stars
and the properties of their planets, we are conducting a survey of
massive subgiants at Keck and Lick Observatories. The decreased
rotation rates and cooler surface temperatures of these evolved
stars 
make them much more ideal Doppler-survey targets compared to their massive
main-sequence progenitors \citep{galland05}. The observed effects of
stellar mass on the properties of planets have important implications
for planet formation modeling \citep{ida05a, kennedy08, kretke09,
  currie09, dr09}; the interpretation of observed structural features in the
disks around massive stars \citep{wyatt99,quillen06,brittain09}; and
the planning of current and future planet  
search efforts, such as the Gemini Planet Imager
\citep[GPI;][]{gpi}, Spectro-Polarimetric High-contrast Exoplanets
REsearch \citep[SPHERE;][]{sphere}, the Near Infrared Coronographic
Imager \citep[NICI;][]{nici}, and {\it Project 1640}  
\citep{p1640}. Our Lick 
survey has resulted in the discovery of 7 new Jovian planets  
orbiting evolved stars more massive than the Sun
\citep{johnson06b,johnson07,johnson08a,peek09,bowler10}. In this
contribution, we 
present the first \npllet\ planets discovered in the expanded Keck
survey.  

\section{A Doppler Survey of Subgiants at Keck Observatory}
\subsection{Target Selection}

We are monitoring the radial velocities (RV) of a sample of 500 evolved
stars at Keck Observatory. The Keck program expands upon our Doppler
survey of 120 subgiants at Lick Observatory, which has been ongoing
since 2004 \citep{johnson06b,peek09}. The stars in the Lick program
have now been folded into the Keck target list and that subset of
brighter subgiants ($V < 7.25$) is currently monitored at both
observatories. We began the Keck survey in 2007 April for
the majority of our target stars, and a handful of stars were part of
the original Keck planet search sample dating as far back as 1997
\citep{marcy08}. 

We selected the targets for the expanded Keck survey from the 
{\it Hipparcos} catalog based on the 
criteria $1.8 < M_V < 3.0$, $0.8 < B-V < 1.1$, and $V \lesssim 8.5$
\citep{hipp, hipp2}. We chose the red cutoff to avoid red giants, the
majority of which are already monitored by other planet search
programs and are known to exhibit velocity jitter $>
10$~\ms\ \citep{sato05,hekker06,nied09}. The  lower $M_V$ restriction 
avoids Cepheid variables, and the upper limit excludes stars with
masses less
than $1.3$~\msun\ when compared to the Solar-metallicity (\feh~$=0$)
stellar model   
tracks of \citet{girardi02}. We also excluded stars in the clump
region ($B-V >  
0.8$, $M_V < 2.0$) to avoid the closely-spaced, and often
overlapping mass tracks in that region of the theoretical H--R
diagram. 

\subsection{Stellar Properties}

Atmospheric parameters of the target stars are estimated from
iodine-free, ``template'' spectra using the LTE
spectroscopic analysis package {\it Spectroscopy Made Easy}
\citep[SME;][]{valenti96}, as described by 
\citet{valenti05} and \citet{fischer05b}. To constrain the low surface
gravities of the evolved stars we used the iterative scheme of
\citet{valenti09}, which ties the SME-derived value of $\log{g}$ to
the gravity inferred from the Yonsei-Yale \citep[Y$^2$;][]{y2} stellar model
grids. The analysis yields a best-fit 
estimate of \teff, \logg, \feh, and \vsini. The properties of the
majority of our targets from Lick and Keck are listed in the fourth
edition of the 
Spectroscopic Properties of Cool Stars Catalog (SPOCS IV.; Johnson et
al. 2010, in prep). We adopt the SME parameter uncertainties
described in the error analysis of \citet{valenti05}. 

The luminosity of each star is estimated from the apparent V-band
magnitude, the bolometric correction \citep{flower96}, and the
parallax from {\it Hipparcos} \citep{hipp2}.  From \teff\ and
luminosity, we determine 
the stellar mass, radius, and an age estimate by associating those
observed properties with a model from the Y$^2$ stellar
interior calculations \citep{y2}. We also measure the chromospheric
emission in the \caii\ line cores 
\citep{wright04b, isaacson09}, providing an $S_\mathrm{HK}$ value on the
Mt. Wilson system, which we convert to \rphk\ as per \cite{Noyes84}.

The stellar properties of the \npllet\ stars presented herein 
are summarized in Table \ref{tab:stars}. 

\section{Observations and Analysis}

\subsection{Keck Spectra and Doppler Analysis}

We obtained spectroscopic observations at Keck Observatory using the
HIRES spectrometer with a resolution of $R \approx 55,000$ with the
B5 decker (0\farcs86 width) and red cross-disperser \citep{vogt94}.
We use the HIRES 
exposure meter to ensure that all observations receive uniform flux
levels independent of atmospheric transparency variations, and to
provide the photon-weighted exposure midpoint which is used for the
barycentric correction. Under nominal atmospheric conditions, a $V=8$
target requires an exposure time of 90 seconds and results in a
signal-to-noise ratio of 190 at 5500\,\AA. 

The spectroscopic observations are made through a
temperature-controlled Pyrex cell containing gaseous iodine, which is
placed at the entrance slit of the spectrometer. The dense set of
narrow molecular lines imprinted on each stellar spectrum from
5000~\AA\ to 
6000~\AA\ provides a robust, simultaneous wavelength calibration for
each observation, as well as information about the shape of the
spectrometer's instrumental response \citep{marcy92b}. Doppler shifts
are measured from each spectrum using the modeling procedure described
by \citet{butler96}. The instrumental uncertainty of each measurement
is estimated based on the weighted standard deviation of the mean
Doppler-shift measured from each of $\approx700$ 2--\AA\ spectral
chunks. In a few instances we made two or more successive observations
of the same star and binned the velocities (in 2\,hr time intervals),
thereby reducing the associated measurement uncertainty. 

\subsection{Photometric Measurements}

We also acquired brightness measurements of five of the seven planetary 
candidate host stars with the T3 0.4~m automatic photometric telescope (APT) 
at Fairborn Observatory.  T3 observed each program star differentially with 
respect to two comparisons stars in the following sequence, termed a group 
observation: {\it K,S,C,V,C,V,C,V,C,S,K}, where $K$ is a check (or secondary 
comparison) star, $C$ is the primary comparison star, $V$ is the program 
(normally a variable) star, and $S$ is a sky reading.  Three $V-C$ and 
two $K-C$ differential magnitudes are computed from each sequence and 
averaged to create group means.  Group mean differential magnitudes with 
internal standard deviations greater than 0.01 mag were rejected to eliminate 
the observations taken under non-photometric conditions.  The surviving 
group means were corrected for differential extinction with nightly 
extinction coefficients, transformed to the Johnson system with yearly-mean 
transformation coefficients, and treated as single observations thereafter.  
The typical precision of a single group-mean observation from T3, as measured 
for pairs of constant stars, is $\sim$0.004--0.005 mag 
\citep[e.g.,][tables 2 \& 3]{henry00c}.  Further information on the operation 
of the T3 APT can be found in \citet{henry95,henry95b} and \citet{eaton03}.

Our photometric observations are useful for eliminating potential false 
positives from the sample of new planets. \citet{queloz01b} and 
\citet{paulson04} have demonstrated how rotational modulation in the 
visibility of starspots on active stars can result in periodic radial 
velocity variations and potentially lead to erroneous planetary 
detections.  Photometric results for the stars in the present sample are 
given in Table~\ref{tab:phot}.  Columns 7--10 give the standard deviations of the $V-C$ 
and $K-C$ differential magnitudes in the $B$ and $V$ passbands with the 
$3\sigma$ outliers removed.  With the exception of HD~206610, all of the 
standard deviations are small and approximately equal to the measurement 
precision of the telescope.  

For HD~206610, the standard deviations of the four data sets $(V-C)_B$, 
$(V-C)_V$, $(K-C)_B$, and $(K-C)_V$ are all larger than 0.01 mag and 
indicate photometric variability.  Periodogram analyses revealed that all 
four data sets have a photometric period of 0.09 day and an amplitude of 
$\sim0.03$ mag.  Thus, the variability must arise from HD~206610's primary 
comparsion star ($C$ = HD~205318), which is included in all four data sets. 
Given its period, amplitude, and early-F spectral class, it is probably 
a new $\delta$~Scuti star.  We computed new differential magnitudes for 
HD~206610 using the check star ($K$) to form the variable minus check 
data sets $(V-K)_B$ and $(V-K)_V$.  The standard deviations of these two 
data sets are 0.0066 and 0.0067 mag, respectively. 

Therefore, all five of the planetary candidate stars in Table~\ref{tab:phot} are
photometrically constant to the approximate limit of the APT observations.  
The measured photometric stability supports the planetary interpretation 
of the radial velocity variations.  The 
two stars that we did not observe photometrically, HD~136418 and 
and HD~181342, have similar masses, colors, and activity levels
(Table~\ref{tab:phot})  
as the five stars we did observe and so are likely to be photometrically 
constant as well.

\subsection{Orbit Analysis}
\label{sec:orbit}

For each star we performed a thorough search of the measured
velocities for
the best-fitting, single-planet Keplerian orbital model using the
partially-linearized, least-squares fitting procedure described in
\citet{wrighthoward} and implemented in the IDL package {\tt
  RVLIN}\footnote{http://exoplanets.org/code/}. Before searching for a
best-fitting solution, we 
increased the measurement uncertainties by including an error
contribution due to stellar "jitter." The jitter accounts for any
unmodeled noise sources intrinsic to the star such as rotational
modulation of surface inhomogeneities and pulsation \citep{saar98,
  wright05,makarov09,lagrange10}.  

\begin{figure}[!h]                                                             
\epsscale{1.1}                                                                 
\plotone{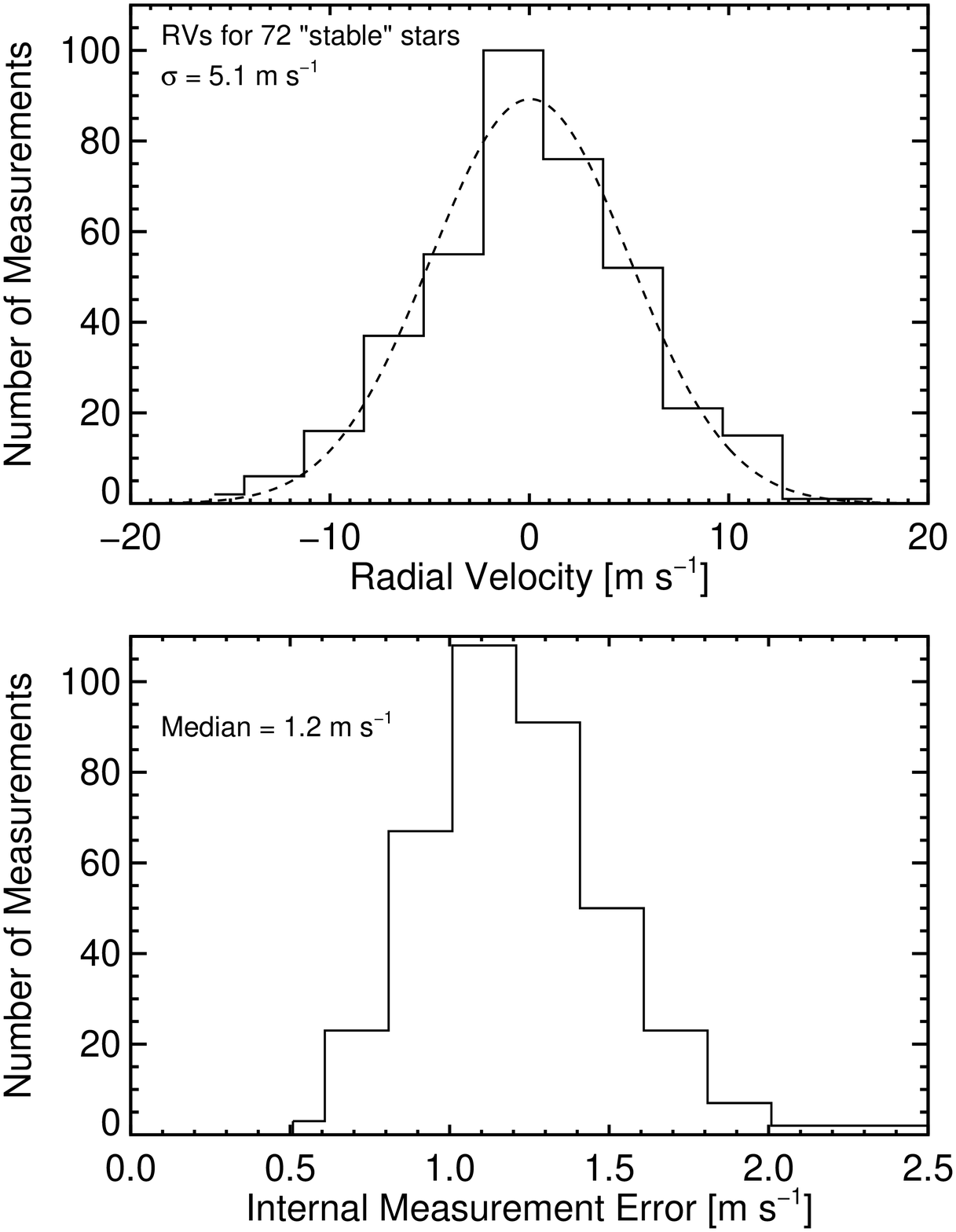}                                               
\caption{ {\it Top}---Distribution of RVs for 72 standard stars,
  comprising a total of 382 measurements. The dashed line shows the
  best-fitting Gaussian with a width $\sigma = 5.1$~\ms. 
{\it Bottom} --Distribution of internal measurement uncertainties for 
  382 RV measurements. The median is 1.2~\ms. Together with the
  distribution of RVs in the top panel, this provides us with a jitter
  estimate of 5~\ms, which we apply to all of the stars presented
  herein.  \label{fig:standard}}                                               
\end{figure}  

We estimate the
jitter for our subgiants based on the velocity variability of a sample
of "stable" stars, for which we have obtained $> 4$ observations over
a time span greater than 2 years. These stars do not show evidence of an
orbital companion, except in a few cases where the stars exhibit a
linear trend. In those cases we remove the trend using a linear fit
and consider the scatter in the residuals. Figure~\ref{fig:standard}
shows the distribution of RVs for all 72 stable stars, comprising 382
measurements. We fit a Gaussian function to the distribution with a
width $\sigma = 5.1$~\ms. The measurement uncertainties, shown in the
lower panel, span
0.6--2.5~\ms, with a median value of 1.2~\ms. We subtracted this 
median internal error in quadrature from the measured width of the
distribution of 
RVs to produce a jitter estimate of 4.95~\ms. In the analysis of each
star's RV time series we round this value up and 
adopt a uniform jitter estimate of 5~\ms, which we add in quadrature
to the measurement uncertainties before searching for the
best--fitting orbit.

After identifying the best--fitting model, we use a Markov-Chain Monte
Carlo (MCMC) algorithm to 
estimate the parameter uncertainties \citep[See,
  e.g.][]{ford05,winn07}.  MCMC is a Bayesian inference 
technique that uses the data to explore the shape of the likelihood
function for each parameter of an input model.  At each step, one 
parameter is selected at random and altered by drawing a random variate
from a normal distribution.  If the resulting $\chi^2$ value for the new
trial orbit is less than the previous $\chi^2$ value, then the trial
orbital parameters are added to the chain.  If not, then the
probability of adopting the new value is set by the ratio of the
probabilities from the previous and current trial steps.  If the
current trial is
rejected then the parameters from the previous step are adopted.  

We alter the standard deviations of the random parameter variates so
that the acceptance rates are between 20\% and 40\%. The initial
parameters are chosen from the best-fitting orbital solutions derived
using the least-squares method described above, and each chain is run
for 10$^7$ steps.  The initial 10\% of the chains are
excluded from the final estimation of parameter uncertainties to
ensure uniform convergence.  We verify that convergence
is reached by running five shorter chains with $10^6$ steps and
checking that the Gelman-Rubin statistic \citep{gelmanrubin} for each
parameter is near unity ($\lesssim$ 1.02) and that a visual inspection
of the history plots suggests stability.

The resulting ``chains'' of
parameters form the posterior probability distribution, 
from which we select the 15.9 and 84.1
percentile levels in the cumulative distributions as the ``one-sigma''
confidence limits. In most cases the posterior probability
distributions were approximately Gaussian.
 
\subsection{Testing RV Trends}

We use the Bayesian Information Criterion \citep[BIC;][]{schwarz78,
  liddle04} and the  MCMC posterior probability density functions (pdf)
to determine whether there is evidence  for a linear velocity trend
\citep{bowler10}.  The BIC rewards better-fitting models but
penalizes overly  complex models, and is given by

\begin{equation}
\mathrm{BIC} \equiv -2 \ln \mathcal{L_\mathrm{max}} + k \ln N,
\end{equation}

\noindent where $\mathcal{L_\mathrm{max}} \propto
exp({-\chi_{min}^2/2})$ is
the maximum likelihood 
for a particular model with $k$ free parameters and $N$ data
points\footnote{The relationship between
  $\mathcal{L_\mathrm{max}}$ and $\chi_{min}^2$ is only valid under
  the assumption that the errors are described by a Gaussian, which
  is approximately valid for our analyses.}. A difference of $\gtrsim$
2 between BIC values with and without a trend  indicates that there is
sufficient evidence for a more complex model \citep{kuha04}. 

We also  use the MCMC-derived pdf for the velocity trend parameter 
to estimate the probability that a trend is actually present in the
data. If the 99.7 percentile of the pdf lies above or below
0~\ms~yr$^{-1}$ then 
we adopt the model with the trend.  The BIC and MCMC methods yield
consistent results for the planet candidates described in
\S~\ref{results}. 

\subsection{False-Alarm Evaluation}

For each planet candidate we consider the null--hypothesis that the
apparent 
periodicity arose by chance from larger-than-expected radial velocity
fluctuations and sparse sampling. We test this possibility by
calculating the false-alarm probability (FAP) based on the
goodness-of-fit statistic $\Delta\chi^2_\nu$  
\citep{howard09,marcy05a,cumming04}, which is the difference between
two values of $\chi^2_\nu$: one from the single-planet Keplerian fit and
one from the fit of a linear trend to the data. Each trial is
constructed by keeping the times of observation fixed and scrambling
the measurements, with replacement.  We record the
$\Delta\chi^2_\nu$  value after each trial and repeat this process for
10,000 trial data sets. For the ensemble set we compare the resulting
distribution of $\Delta\chi^2_\nu$ to the value from the fit to the
original data. The planets presented below all have FAP~$ < 0.001$,
corresponding to $<0.5$ false alarms for our
sample of 500 stars. 

\section{Results}
\label{results}

\begin{figure}[!t]
\epsscale{1.1}
\plotone{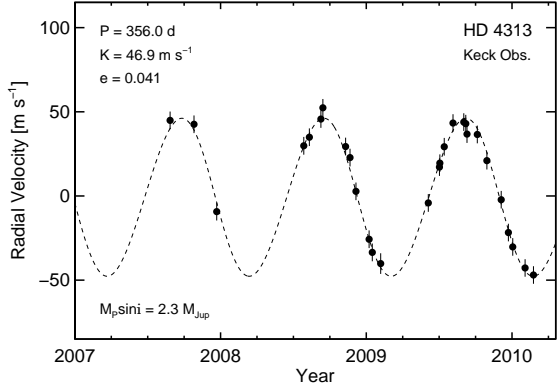}
\caption{Relative RVs of \starB\ measured at Keck Observatory. The error
  bars are the quadrature sum of the internal measurement uncertainties
  and 5~\ms\ of jitter. The dashed line shows the best-fitting orbit
  solution of a single Keplerian orbit. The solution results in
  residuals with an rms scatter of \rmsB~\ms\ and
  \chisq~$=$~\chisB.  \label{fig:hd4313}}
\end{figure}

\begin{figure}[!t]
\epsscale{1.1}
\plotone{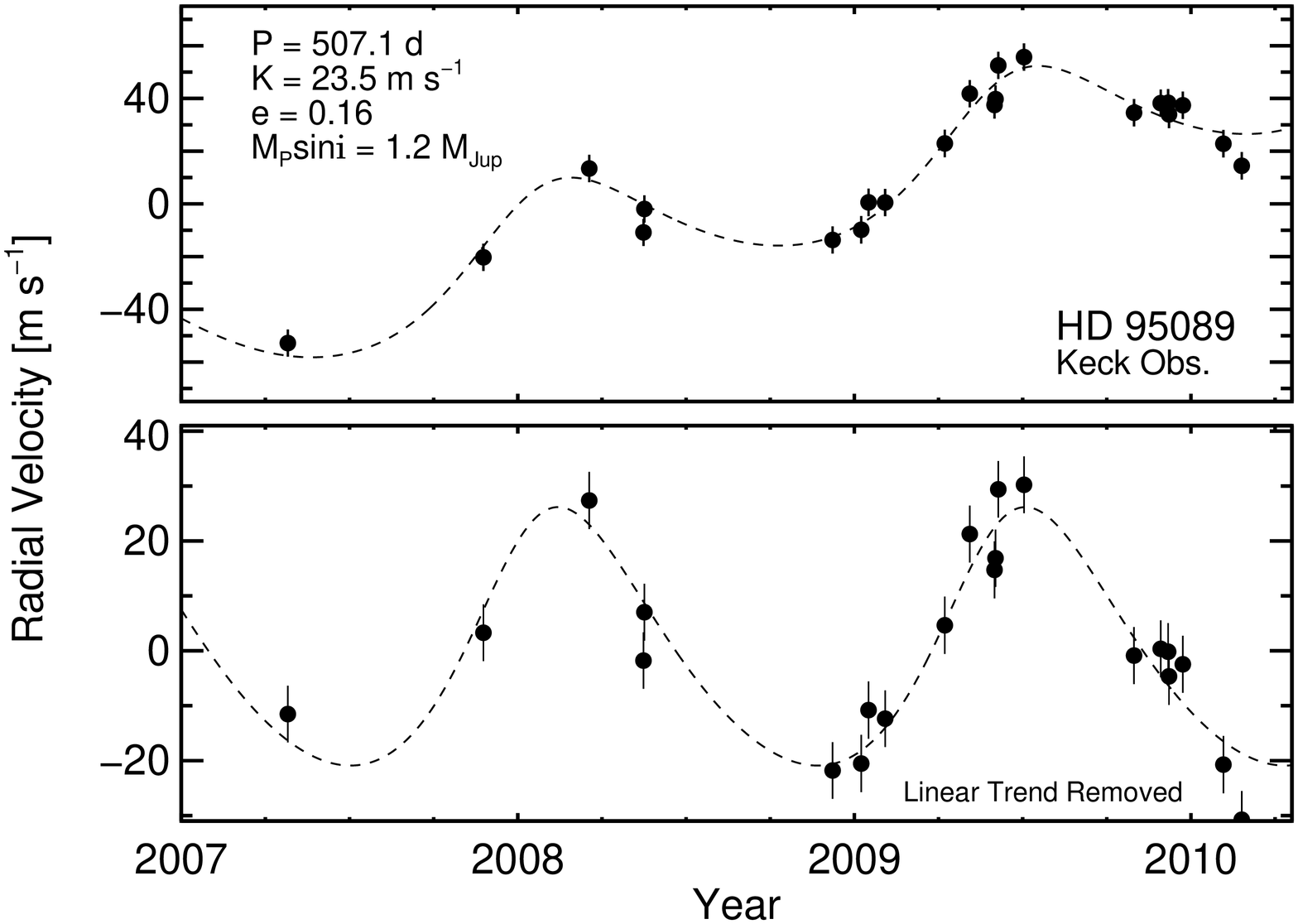}
\caption{Relative RVs of \starC\ measured at Keck Observatory. The error
  bars are the quadrature sum of the internal measurement uncertainties
  and 5~\ms\ of jitter. The dashed line shows the best-fitting orbit
  solution of a single Keplerian orbit plus a linear trend ($dv/dt =
  \trendC \pm \trendeC$~\msy). The solution results in
  residuals with an rms scatter of \rmsC~\ms\ and
  \chisq~$=$~\chisC. he lower panel shows the RVs with the linear
  trend removed.  \label{fig:hd95089}}
\end{figure}

\begin{figure}[!h]
\epsscale{1.1}
\plotone{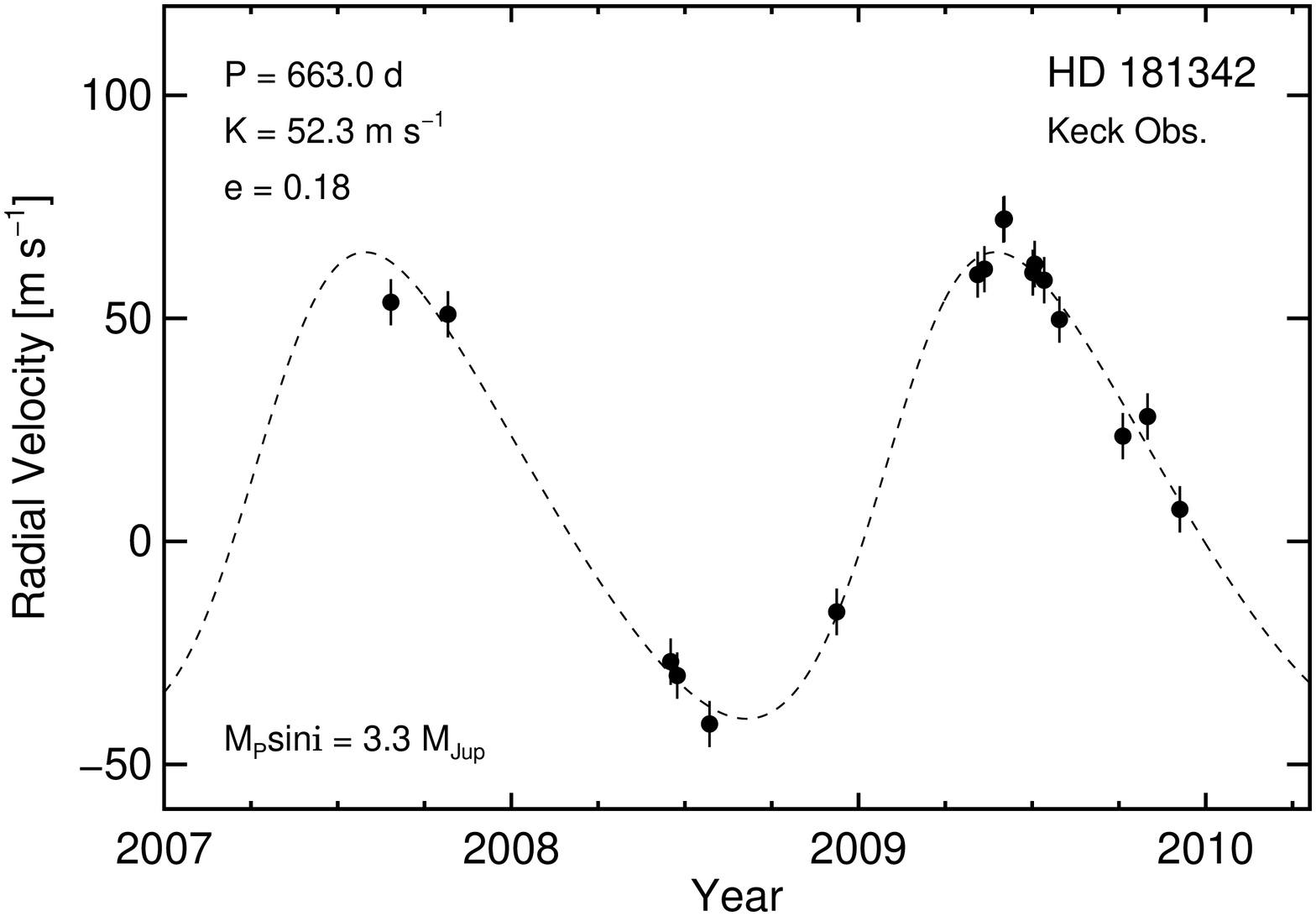}
\caption{Relative RVs of \starD\ measured at Keck
  Observatory. The error 
  bars are the quadrature sum of the internal measurement uncertainties
  and 5~\ms\ of jitter. The dashed line shows the best-fitting orbit
  solution of a single Keplerian orbit. The solution results in
  residuals with an rms scatter of \rmsD~\ms\ and
  \chisq~$=$~\chisD.  \label{fig:hd181342}} 
\end{figure}

\begin{figure}[!h]
\epsscale{1.1}
\plotone{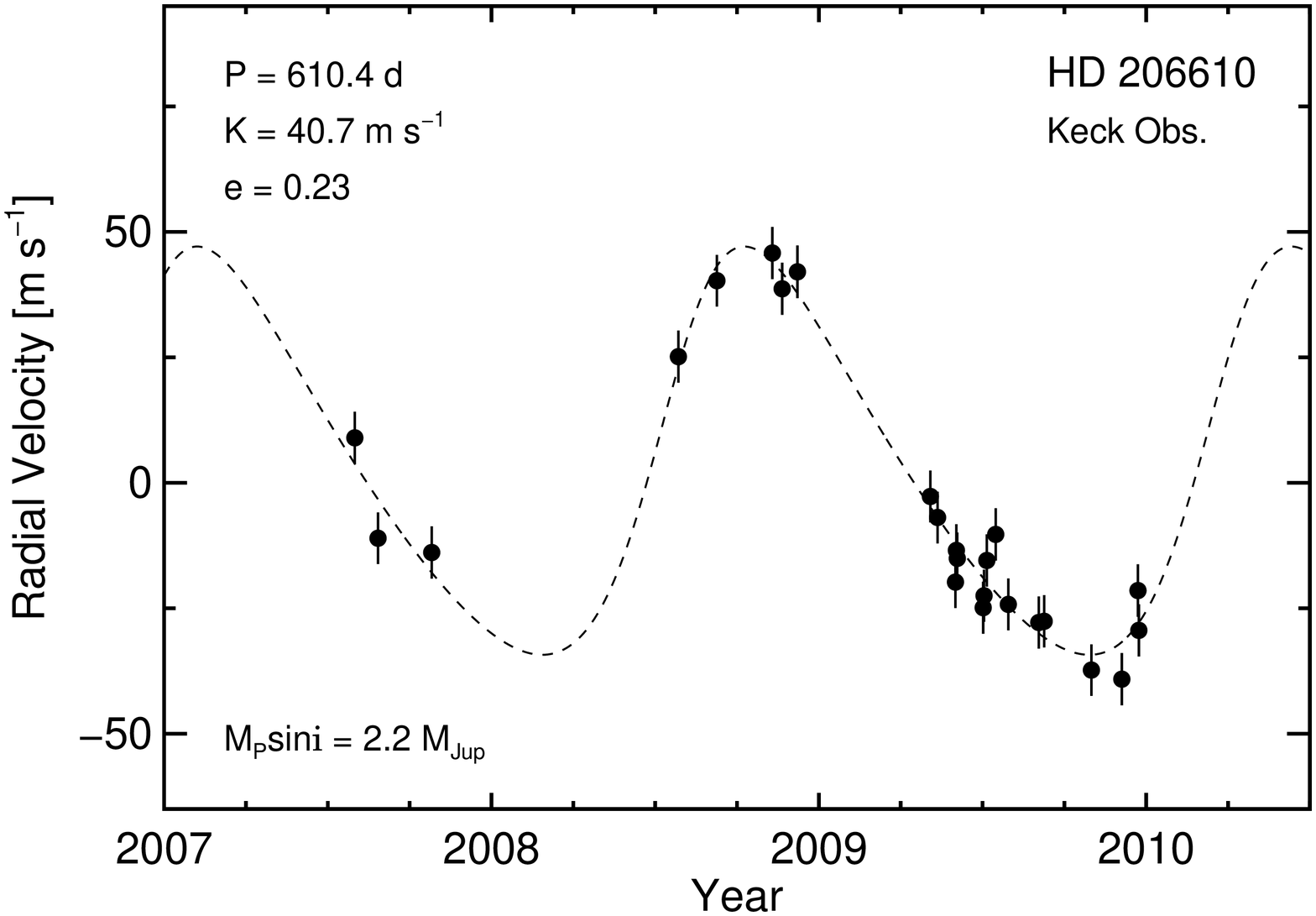}
\caption{Relative RVs of \starE\ measured at Keck Observatory. The error
  bars are the quadrature sum of the internal measurement uncertainties
  and 5~\ms\ of jitter. The dashed line shows the best-fitting orbit
  solution of a single Keplerian orbit. The solution results in
  residuals with an rms scatter of \rmsE~\ms\ and
  \chisq~$=$~\chisE.  \label{fig:hd206610}}
\end{figure}

\begin{figure}[!h]
\epsscale{1.1}
\plotone{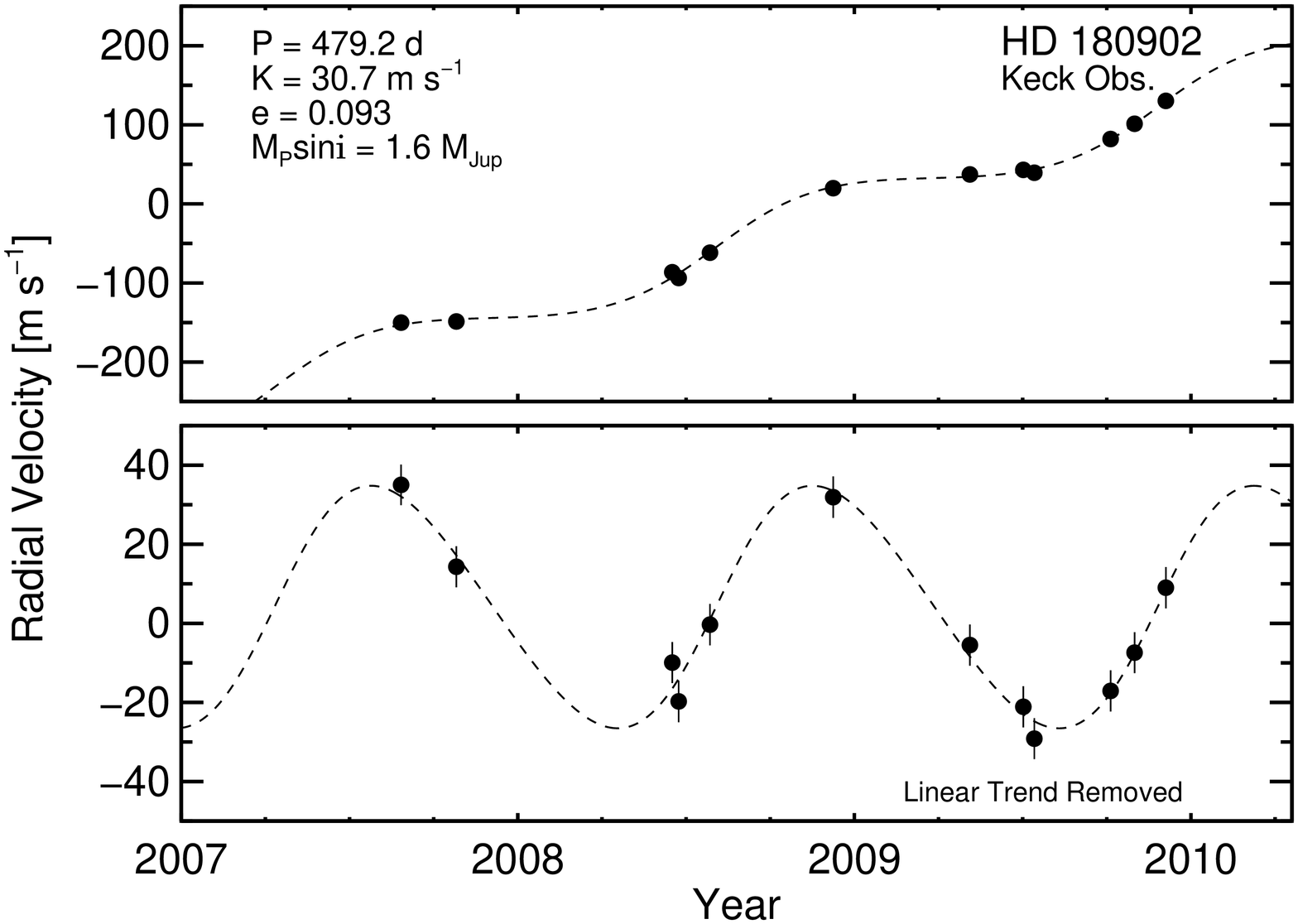}
\caption{Relative RVs of \starH\ measured at Keck Observatory. The error
  bars are the quadrature sum of the internal measurement uncertainties
  and 5~\ms\ of jitter. The dashed line shows the best-fitting orbit
  solution of a single Keplerian orbit plus a linear trend ($dv/dt =
  \trendH \pm \trendeH$~\msy). The solution results in
  residuals with an rms scatter of \rmsH~\ms\ and
  \chisq~$=$~\chisH. The lower panel shows the RVs with the linear
  trend removed. \label{fig:hd180902}}
\end{figure}

\begin{figure}[!h]
\epsscale{1.1}
\plotone{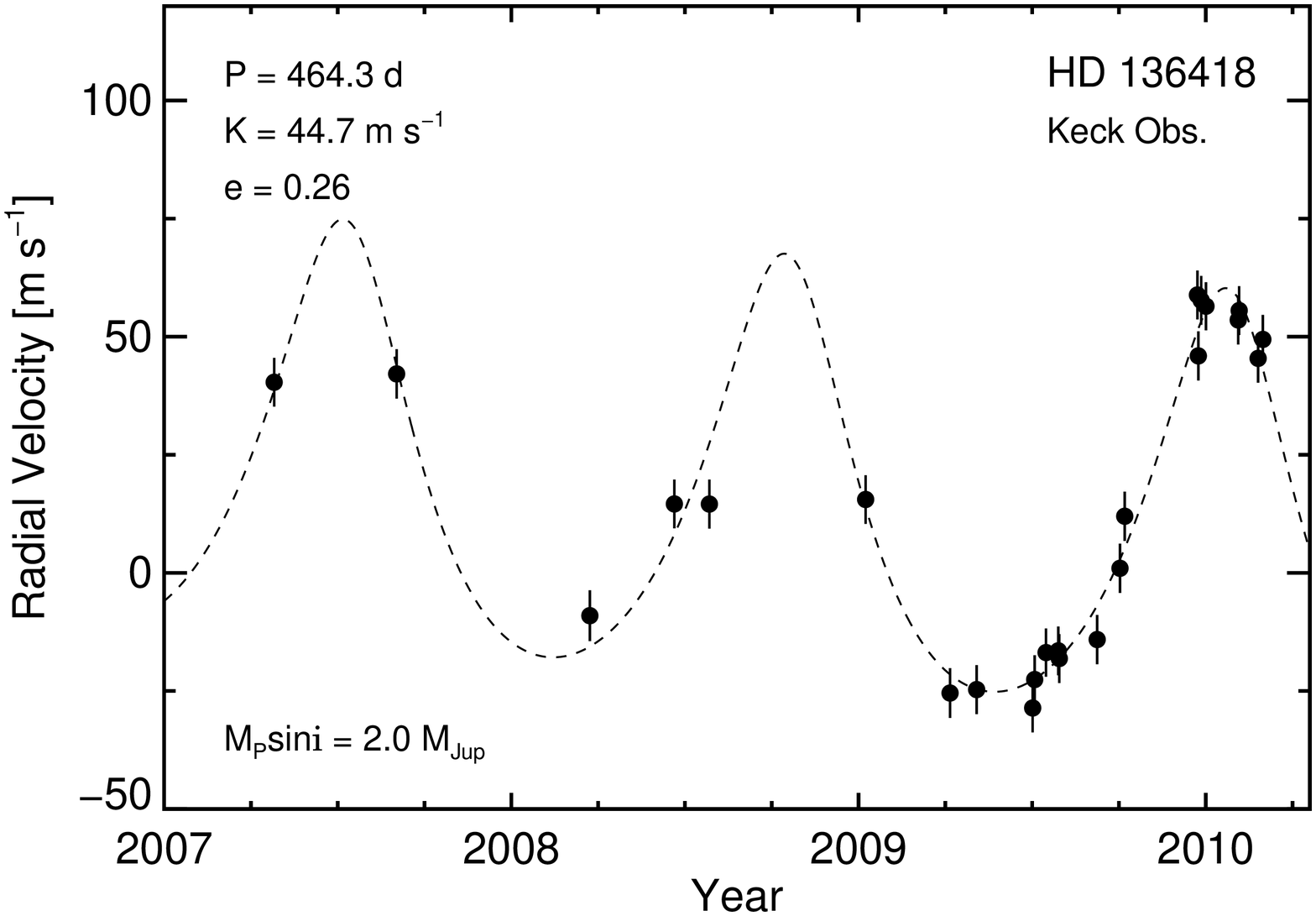}
\caption{Relative RVs of \starF\ measured at Keck Observatory. The error
  bars are the quadrature sum of the internal measurement uncertainties
  and 5~\ms\ of jitter. The dashed line shows the best-fitting orbit
  solution of a single Keplerian orbit plus a linear trend ($dv/dt =
  \trendF \pm \trendeF$~\msy). The solution results in
  residuals with an rms scatter of \rmsF~\ms\ and
  \chisq~$=$~\chisF. \starF\ has a mass $M_\star = $~\mstarF~\msun,
  making it a former F-type star. \label{fig:hd135418}}
\end{figure}

\begin{figure}[!h]
\epsscale{1.1}
\plotone{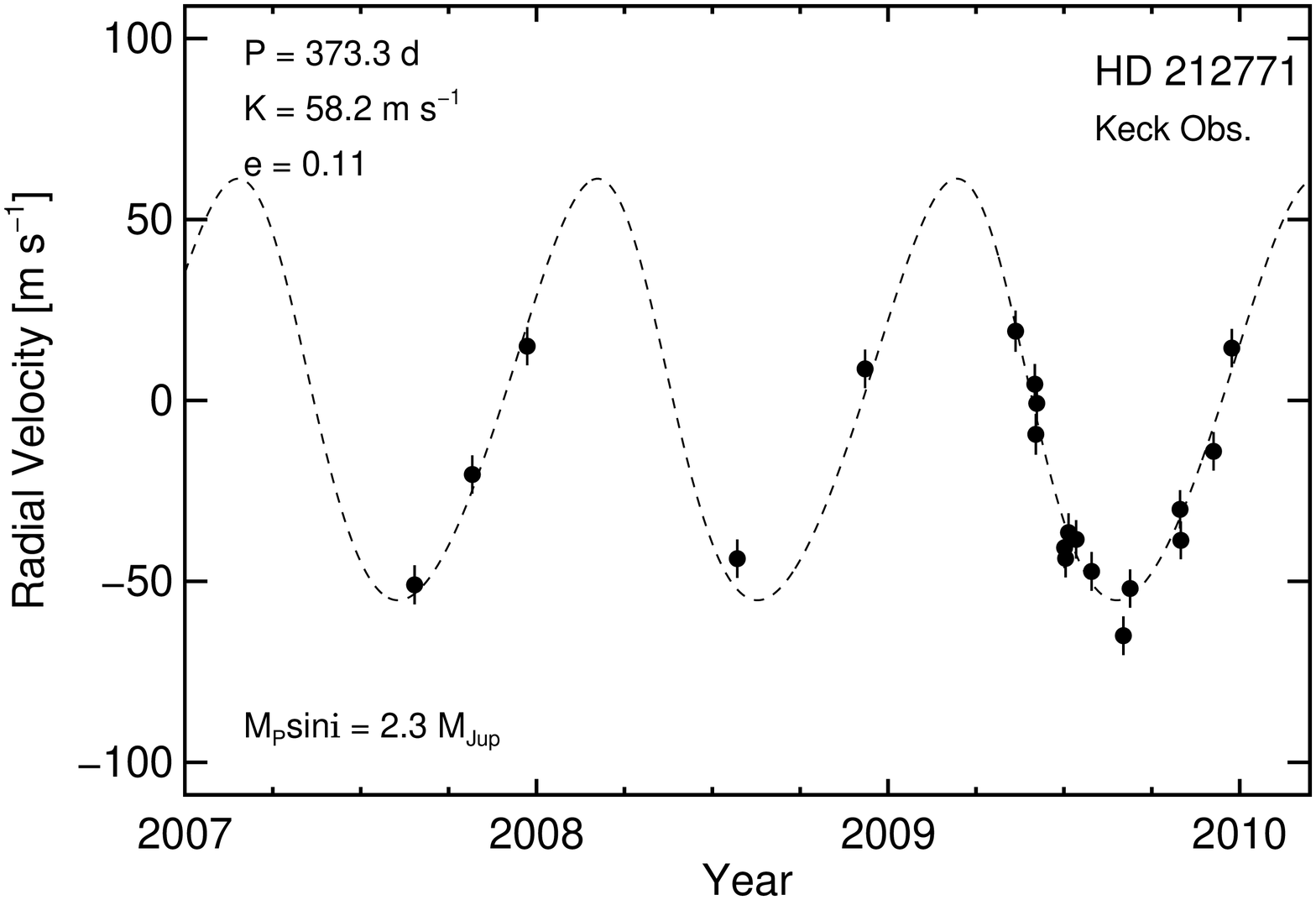}
\caption{Relative radial velocity measurements of \starG. The error
  bars are the quadrature sum of the internal measurement uncertainties
  and 5~\ms\ of jitter. The dashed line shows the best-fitting orbit
  solution of a single Keplerian orbit. The solution results in
  residuals with an rms scatter of \rmsG~\ms\ and
  \chisq~$=$~\chisG. \starG\ has a mass $M_\star = $~\mstarG~\msun,
  indicating that it was either an early-G or late-F star when it was
  on the main sequence. \label{fig:hd212771}}
\end{figure}

We have detected \npllet\ new Jovian planets orbiting evolved,
subgiant stars. The RV time series of each host-star is plotted in
Figures~\ref{fig:hd4313}--\ref{fig:hd212771}, where the error bars
show the quadrature sum of the internal errors and the jitter estimate
of 5~\ms, as described in \S~\ref{sec:orbit}. The RV measurements for
each star are listed in Tables~\ref{vel4313}--\ref{vel212771},
together with the Julian Date of observation and the internal
measurement uncertainties (without jitter). The best--fitting orbital
parameters and physical characteristics of the planets are summarized
in Table~\ref{tab:planets}, along with their uncertainties. When
appropriate we list notes for some of the individual planetary
systems. 

{\it \starC, \starF, \starH}---The orbit models for these three
stars include linear trends. We interpret the linear trend as a second
orbital companion with a period longer than the time baseline of the
observations. 

{\it \starD}---The time sampling of
\starD\ is sparser than most of the the other stars presented
in this work. However, the large amplitude of the variations and
observations clustered near the quadrature points result in a
well-defined \chis\ minimum in the orbital parameter space. The FAP
for the orbit solution is $0.0064$\%.

{\it \starG}---This low-mass subgiant has a mass $M_\star =
\mstarG$~\msun, indicating that it had a spectral type of  early-G to
late-F while on the main sequence. In addition to being one of our
least massive targets, \starG\ is also one of the most
metal-poor stars in the Keck sample \feh~$=\feG$. 

\section{Summary and Discussion}

We report the detection of \npllet\ new Jovian planets orbiting 
evolved stars. These detections come from the sample of
subgiants that we are monitoring at Lick and Keck
Observatories. The host-stars have masses in the range
$\mstarG$~\msun\ to 1.9~\msun, radii $\rstarF \leq
R_\star/R_\odot \leq \rstarE$, and metallicities $-0.21
\leq$~\feh~$ \leq +0.26$. Five of the host-stars have 
masses $M_\star > 1.5$~\msun, and are therefore the evolved
counterparts of the A-type stars. We also derived a jitter estimate
for our sample of evolved stars and 
find that subgiants are typically stable to within 5~\ms. The
observed jitter of subgiants makes them uniquely stable Doppler
targets among massive, evolved stars \citep{fischer03,hekker06}. 

\citet{bowler10} found that the minimum masses and
semimajor axes of planets 
around A stars are very different from those of planets around
FGK stars. Their findings suggest that the formation and migration
mechanisms of planets changes dramatically with increasing stellar
mass. The planets reported in this work strengthen that conclusion. 
The five new planets we have discovered around stars with $M_\star >
1.5$~\msun\ all orbit 
beyond 1~AU and have minimum masses \msini~$> 
1$~\mjup. These properties contrast with those of planets orbiting
less massive stars, which have a nearly flat distribution in $\log{a}$
from $a = 0.05$~AU to $a = 1$~AU \citep{cumming08}, and a steeply
rising mass function with $d\ln{N}/d\ln{M_P} = -1.4$.  
Successful theories of the origin and orbital evolution of giant
planets will need 
to account for the discontinuity between the distributions of orbital
parameters for planets around Sun-like and A-type stars
\citep{kennedy08,currie09,kretke09}. 

The abundance of super-Jupiters (\msini~$ > 1$~\mjup) detected
around massive stars bodes well for future direct-imaging surveys. In
addition to harboring the massive planets that are predicted to be
the most easily detectable in high-contrast images, A-type
dwarfs have the added benefit of being naturally young. A
2~\msun\ star has a main-sequence lifetime of only $\sim1$~Gyr, which
means 
that Jovian planets in wide orbits will be young and 
thermally bright. Three of the planets in Table~\ref{tab:planets} show
linear velocity trends indicative of additional long-period
companions. These linear trends provide clear markers of massive
objects in wide orbits around nearby stars, and therefore warrant
additional scrutiny from RV monitoring and high-contrast imaging.

\acknowledgments
We thank the many observers who contributed to the observations
reported here. We gratefully acknowledge the efforts and dedication
of the Keck Observatory staff, especially Grant Hill, Scott Dahm and
Hien Tran for their support of 
HIRES and Greg Wirth for support of remote observing. We are also
grateful to the time assignment committees of NASA, NOAO, Caltech, and
the University of California for their generous allocations of
observing time. A.\,W.\,H.\ gratefully acknowledges support from a
Townes Post-doctoral Fellowship at the U.\,C.\ Berkeley Space Sciences 
Laboratory. J.\,A.\,J.\ thanks the NSF Astronomy and Astrophysics
Postdoctoral Fellowship program for support in the years leading to
the completion of this work, and acknowledges support form NSF grant
AST-0702821 and the NASA Exoplanets Science Institute
(NExScI). G.\,W.\,M.\ acknowledges NASA grant NNX06AH52G.   
J.\,T.\,W.\ received support from NSF grant AST-0504874.
G.\,W.\,H acknowledges support from NASA, NSF,
Tennessee State University, and the State of Tennessee through its
Centers of Excellence program. 
Finally, the authors wish to extend special thanks to those of
Hawaiian ancestry  on whose sacred mountain of Mauna Kea we are
privileged to be guests.   
Without their generous hospitality, the Keck observations presented herein
would not have been possible.

\bibliography{}
\clearpage

\begin{deluxetable}{lllllllll}
\tablecaption{Stellar Parameters\label{tab:stars}}
\tablewidth{0pt}
\tablehead{
  \colhead{Parameter} & 
  \colhead{\starB}    & % 4313
  \colhead{\starC}    & % 95089
  \colhead{\starD}    & % 181342
  \colhead{\starE}    & % 206610
  \colhead{\starH}    & % 180902
  \colhead{\starF}    & % 136418
  \colhead{\starG}    \\ % 212771
}
\startdata
V             & 
   \vmagB        &
   \vmagC        & 
   \vmagD        & 
   \vmagE        & 
   \vmagH        & 
   \vmagF        & 
   \vmagG        \\
$B-V$ &
   \bvB         & 
   \bvC         & 
   \bvD         & 
   \bvE         & 
   \bvH         & 
   \bvF         & 
   \bvG         \\ 
Distance (pc)   &
   \dB~(\deB)          & 
   \dC~(\deC)          & 
   \dD~(\deD)          & 
   \dE~(\deE)          & 
   \dH~(\deH)          & 
   \dF~(\deF)          & 
   \dG~(\deG)          \\
$M_V$ &
   \mvB~(\mveB)         & 
   \mvC~(\mveC)         & 
   \mvD~(\mveD)         & 
   \mvE~(\mveE)         & 
   \mvH~(\mveH)         & 
   \mvF~(\mveF)         & 
   \mvG~(\mveG)         \\ 
${\rm [Fe/H]}$  &
   \feB~(0.03)         &       
   \feC~(0.03)         &       
   \feD~(0.03)         &       
   \feE~(0.03)         &       
   \feH~(0.03)         &       
   \feF~(0.03)         &       
   \feG~(0.03)         \\
$T_{\rm eff}$~(K)   &
   \teffB~(44)       &
   \teffC~(44)       &
   \teffD~(44)       &
   \teffE~(44)       &
   \teffH~(44)       &
   \teffF~(44)       &
   \teffG~(44)       \\
\vsini~(\ks)    &
   \vsiniB~(0.5)      & 
   \vsiniC~(0.5)      & 
   \vsiniD~(0.5)      & 
   \vsiniE~(0.5)      & 
   \vsiniH~(0.5)      & 
   \vsiniF~(0.5)      & 
   \vsiniG~(0.5)      \\
$\log{g}$       &
   \loggB~(0.06)        &
   \loggC~(0.06)        &
   \loggD~(0.06)        &
   \loggE~(0.06)        &
   \loggH~(0.06)        &
   \loggF~(0.06)        &
   \loggG~(0.06)        \\
$M_{*}$~(\msun) &
   \mstarB~(0.12)      &
   \mstarC~(0.11)      &
   \mstarD~(0.13)      &
   \mstarE~(0.11)      &
   \mstarH~(0.11)      &
   \mstarF~(0.09)      &
   \mstarG~(0.08)      \\
$R_{*}$~(\rsun) &
   \rstarB~(\rstareB)      &
   \rstarC~(\rstareC)      &
   \rstarD~(\rstareD)      &
   \rstarE~(\rstareE)      &
   \rstarH~(\rstareH)      &
   \rstarF~(\rstareF)      &
   \rstarG~(\rstareG)      \\
$L_{*}$~(\lsun) &
   \lstarB~(\lstareB)      &
   \lstarC~(\lstareC)      &
   \lstarD~(\lstareD)      &
   \lstarE~(\lstareE)      &
   \lstarH~(\lstareH)      &
   \lstarF~(\lstareF)      &
   \lstarG~(\lstareG)      \\
Age~(Gyr)       &
   \ageB~(\ageeB)           &
   \ageC~(\ageeC)           &
   \ageD~(\ageeD)           &
   \ageE~(\ageeE)           &
   \ageH~(\ageeH)           &
   \ageF~(\ageeF)           &
   \ageG~(\ageeG)           \\
$S_{HK}$        &
   \sB      &      
   \sC      &      
   \sD      &      
   \sE      &      
   \sH      &      
   \sF      &      
   \sG      \\     
$\log R'_{HK}$  &
   \rphkB    &
   \rphkC    &
   \rphkD    &
   \rphkE    &
   \rphkH    &
   \rphkF    &
   \rphkG    \\
\enddata
\end{deluxetable}

\tabletypesize{\tiny}
\begin{deluxetable}{lllcccccccc}
\tablewidth{0pt}
\tablecaption{Summary of Photometric Observations From Fairborn Observatory\label{tab:phot}}
\tablehead{
\colhead{Program} & \colhead{Comparison} & \colhead{Check} & 
\colhead{Date Range} & \colhead{Duration} & \colhead{} & 
\colhead{$\sigma{(V-C)}_B$} & \colhead{$\sigma{(V-C)}_V$} &
\colhead{$\sigma{(K-C)}_B$} & \colhead{$\sigma{(K-C)}_V$} & \colhead{} \\
\colhead{Star} & \colhead{Star} & \colhead{Star} 
& \colhead{(HJD $-$ 2,440,000)} & \colhead{(days)} & \colhead{$N_{obs}$} & 
\colhead{(mag)} & \colhead{(mag)} & \colhead{(mag)} & \colhead{(mag)} &
\colhead{Variability} \\
\colhead{(1)} & \colhead{(2)} & \colhead{(3)} & \colhead{(4)} & \colhead{(5)} &
\colhead{(6)} & \colhead{(7)} & \colhead{(8)} & \colhead{(9)} & \colhead{(10)} &
\colhead{(11)}
}
\startdata
 HD 4313 &   HD 4627 &   HD 4526 & 54756--55222 & 466 & 188 & 0.0039 & 0.0046 & 0.0038 & 0.0056 & Constant \\
 HD 95089 &  HD 94401 &  HD 93102 & 55139--55237 &  98 &  58 & 0.0056 & 0.0059 & 0.0045 & 0.0064 & Constant \\
HD 180902 & HD 179949 & HD 181240 & 55122--55126 &   4 &   3 & 0.0032 & 0.0031 & 0.0010 & 0.0056 & Constant? \\
HD 206610 & HD 205318 & HD 208703 & 54756--55170 & 414 &  88 & 0.0183\tablenotemark{a} & 0.0140\tablenotemark{a} & 0.0172 & 0.0124 & Constant \\
HD 212771 & HD 212270 & HD 213198 & 55119--55170 &  51 &  60 & 0.0043 & 0.0048 & 0.0051 & 0.0065 & Constant \\
\enddata
\tablenotetext{a}{Comparison star HD 205318 is variable in brightness, so we 
recomputed the standard deviations of the program star from the $(V-K)_B$
and $(V-K)_V$ differential magnitudes and find them to be 0.0066 and
0.0067 mag in B and V, respectively.} 
\tabletypesize{\scriptsize}
\end{deluxetable}

\begin{deluxetable}{lll}
\tablecaption{Radial Velocities for HD 4313\label{vel4313}}
\tablewidth{0pt}
\tablehead{
\colhead{JD} &
\colhead{RV} &
\colhead{Uncertainty} \\
\colhead{-2440000} &
\colhead{(m~s$^{-1}$)} &
\colhead{(m~s$^{-1}$)} 
}
\startdata
14339.932 &   23.92 &  1.57 \\
14399.842 &   21.61 &  1.59 \\
14456.806 &  -30.29 &  1.61 \\
14675.006 &    8.82 &  1.71 \\
14689.004 &   13.91 &  1.60 \\
14717.945 &   24.71 &  1.53 \\
14722.895 &   31.39 &  1.60 \\
14779.854 &    8.37 &  1.70 \\
14790.889 &    1.82 &  1.64 \\
14805.807 &  -18.24 &  1.54 \\
14838.768 &  -46.70 &  1.61 \\
14846.745 &  -54.50 &  1.69 \\
14867.754 &  -61.16 &  3.56 \\
14987.118 &  -25.11 &  1.69 \\
15015.049 &   -3.83 &  1.54 \\
15016.081 &   -1.47 &  1.41 \\
15027.089 &    8.26 &  1.61 \\
15049.038 &   22.34 &  1.56 \\
15076.091 &   22.90 &  1.60 \\
15081.091 &   21.97 &  1.55 \\
15084.143 &   15.79 &  1.60 \\
15109.955 &   15.55 &  1.63 \\
15133.975 &    0.00 &  1.60 \\
15169.860 &  -23.21 &  1.57 \\
15187.855 &  -42.72 &  1.55 \\
15198.771 &  -51.25 &  1.55 \\
15229.722 &  -63.75 &  1.43 \\
15250.713 &  -67.93 &  1.50
\\
\enddata
\end{deluxetable}

\begin{deluxetable}{lll}
\tablecaption{Radial Velocities for HD 95089\label{vel95089}}
\tablewidth{0pt}
\tablehead{
\colhead{JD} &
\colhead{RV} &
\colhead{Uncertainty} \\
\colhead{-2440000} &
\colhead{(m~s$^{-1}$)} &
\colhead{(m~s$^{-1}$)} 
}
\startdata
14216.851 &  -11.35 &  1.26 \\
14429.126 &    3.46 &  1.35 \\
14543.962 &   27.54 &  1.44 \\
14602.807 &   -1.61 &  1.19 \\
14603.793 &    7.19 &  1.39 \\
14808.056 &  -21.62 &  1.25 \\
14839.102 &  -20.36 &  1.49 \\
14847.054 &  -10.63 &  1.52 \\
14865.063 &  -12.19 &  1.27 \\
14929.816 &    4.82 &  1.47 \\
14956.912 &   21.43 &  1.36 \\
14983.770 &   14.90 &  1.40 \\
14984.829 &   17.03 &  1.50 \\
14987.835 &   29.57 &  1.28 \\
15015.761 &   30.40 &  1.33 \\
15135.149 &   -0.70 &  1.42 \\
15164.109 &    0.53 &  1.34 \\
15172.146 &    0.00 &  1.45 \\
15173.095 &   -4.48 &  1.47 \\
15188.103 &   -2.29 &  1.42 \\
15232.135 &  -20.55 &  1.49 \\
15252.034 &  -30.59 &  1.53
\\
\enddata
\end{deluxetable}

\begin{deluxetable}{lll}
\tablecaption{Radial Velocities for HD 181342\label{vel181342}}
\tablewidth{0pt}
\tablehead{
\colhead{JD} &
\colhead{RV} &
\colhead{Uncertainty} \\
\colhead{-2440000} &
\colhead{(m~s$^{-1}$)} &
\colhead{(m~s$^{-1}$)} 
}
\startdata
14339.768 &    2.69 &  1.28 \\
14399.741 &    0.00 &  1.33 \\
14634.063 &  -77.87 &  1.43 \\
14641.000 &  -81.00 &  1.44 \\
14674.973 &  -91.85 &  1.32 \\
14808.687 &  -66.72 &  1.50 \\
14957.027 &    8.89 &  1.28 \\
14964.119 &   10.11 &  1.29 \\
14984.083 &   21.20 &  1.31 \\
14985.112 &   21.37 &  1.35 \\
15015.014 &    9.33 &  1.22 \\
15016.963 &   11.24 &  1.41 \\
15026.967 &    7.63 &  1.38 \\
15042.963 &   -1.18 &  1.38 \\
15109.749 &  -27.32 &  1.34 \\
15135.743 &  -22.92 &  1.38 \\
15169.686 &  -43.74 &  1.42
\\
\enddata
\end{deluxetable}

\begin{deluxetable}{lll}
\tablecaption{Radial Velocities for HD 206610\label{vel206610}}
\tablewidth{0pt}
\tablehead{
\colhead{JD} &
\colhead{RV} &
\colhead{Uncertainty} \\
\colhead{-2440000} &
\colhead{(m~s$^{-1}$)} &
\colhead{(m~s$^{-1}$)} 
}
\startdata
14313.980 &   22.85 &  1.48 \\
14339.843 &    2.84 &  1.25 \\
14399.759 &    0.00 &  1.48 \\
14674.964 &   39.03 &  1.42 \\
14717.919 &   54.16 &  1.20 \\
14779.822 &   59.68 &  1.36 \\
14790.746 &   52.55 &  1.42 \\
14807.783 &   55.93 &  1.57 \\
14956.102 &   11.16 &  1.28 \\
14964.118 &    6.98 &  1.26 \\
14984.082 &   -5.90 &  1.27 \\
14985.111 &    0.42 &  1.43 \\
14986.113 &   -1.17 &  1.27 \\
15015.022 &  -11.00 &  1.36 \\
15015.956 &   -8.62 &  1.37 \\
15019.066 &   -1.59 &  1.47 \\
15029.011 &    3.61 &  1.46 \\
15043.058 &  -10.34 &  1.31 \\
15077.058 &  -13.95 &  1.40 \\
15083.055 &  -13.70 &  1.41 \\
15135.761 &  -23.43 &  1.13 \\
15169.700 &  -25.24 &  1.46 \\
15187.696 &   -7.58 &  1.50 \\
15188.691 &  -15.51 &  1.40
\\
\enddata
\end{deluxetable}

\begin{deluxetable}{lll}
\tablecaption{Radial Velocities for HD 180902\label{vel180902}}
\tablewidth{0pt}
\tablehead{
\colhead{JD} &
\colhead{RV} &
\colhead{Uncertainty} \\
\colhead{-2440000} &
\colhead{(m~s$^{-1}$)} &
\colhead{(m~s$^{-1}$)} 
}
\startdata
14339.767 &   40.51 &  1.17 \\
14399.740 &   19.80 &  1.37 \\
14634.062 &   -4.44 &  1.37 \\
14640.998 &  -14.25 &  1.59 \\
14674.972 &    5.16 &  1.51 \\
14808.686 &   37.39 &  1.56 \\
14957.026 &    0.00 &  1.44 \\
15015.013 &  -15.62 &  1.39 \\
15026.965 &  -23.67 &  1.33 \\
15109.748 &  -11.60 &  1.45 \\
15135.741 &   -1.92 &  1.33 \\
15169.685 &   14.50 &  1.51
\\
\enddata
\end{deluxetable}

\begin{deluxetable}{lll}
\tablecaption{Radial Velocities for HD 136418\label{vel136418}}
\tablewidth{0pt}
\tablehead{
\colhead{JD} &
\colhead{RV} &
\colhead{Uncertainty} \\
\colhead{-2440000} &
\colhead{(m~s$^{-1}$)} &
\colhead{(m~s$^{-1}$)} 
}
\startdata
14216.954 &   25.79 &  1.27 \\
14345.803 &   27.56 &  1.50 \\
14549.041 &  -23.63 &  1.97 \\
14637.956 &    0.03 &  1.26 \\
14674.797 &    0.00 &  1.32 \\
14839.165 &    0.97 &  1.21 \\
14927.948 &  -40.01 &  1.66 \\
14955.930 &  -39.28 &  1.38 \\
15014.866 &  -43.17 &  1.21 \\
15016.982 &  -37.15 &  1.13 \\
15028.959 &  -31.44 &  0.96 \\
15041.838 &  -31.07 &  1.26 \\
15042.875 &  -32.70 &  1.33 \\
15082.729 &  -28.68 &  1.46 \\
15106.719 &  -13.60 &  1.43 \\
15111.702 &   -2.58 &  1.49 \\
15188.168 &   44.27 &  1.27 \\
15189.147 &   31.37 &  1.37 \\
15192.148 &   43.12 &  1.31 \\
15197.173 &   41.88 &  0.99 \\
15231.100 &   38.97 &  1.41 \\
15232.083 &   40.97 &  1.32 \\
15252.046 &   30.85 &  1.26 \\
15256.997 &   34.86 &  1.34
\\
\enddata
\end{deluxetable}

\begin{deluxetable}{lll}
\tablecaption{Radial Velocities for HD 212771\label{vel212771}}
\tablewidth{0pt}
\tablehead{
\colhead{JD} &
\colhead{RV} &
\colhead{Uncertainty} \\
\colhead{-2440000} &
\colhead{(m~s$^{-1}$)} &
\colhead{(m~s$^{-1}$)} 
}
\startdata
14339.830 &  -20.86 &  2.03 \\
14399.773 &    9.64 &  1.79 \\
14456.785 &   45.09 &  1.68 \\
14675.025 &  -13.62 &  1.77 \\
14807.780 &   38.82 &  1.84 \\
14964.124 &   49.23 &  2.70 \\
14984.088 &   34.58 &  2.51 \\
14985.116 &   20.73 &  2.62 \\
14986.117 &   29.29 &  2.77 \\
15015.023 &  -10.59 &  2.63 \\
15016.078 &  -13.55 &  1.63 \\
15019.080 &   -6.43 &  1.88 \\
15027.008 &   -8.30 &  1.80 \\
15043.060 &  -17.15 &  1.96 \\
15076.073 &  -34.91 &  1.91 \\
15083.064 &  -21.87 &  1.74 \\
15134.930 &    0.00 &  1.74 \\
15135.765 &   -8.56 &  1.52 \\
15169.702 &   16.03 &  1.77 \\
15188.695 &   44.57 &  1.77
\\
\enddata
\end{deluxetable}

\tabletypesize{\scriptsize}
\begin{deluxetable}{lllllllll}
\tablecaption{Orbital Parameters\label{tab:planets}}
\tablewidth{0pt}
\tablehead{\colhead{Parameter} &
  \colhead{HD\,4313\,b}    &
  \colhead{HD\,95089\,b}    &
  \colhead{HD\,181342\,b}    &
  \colhead{HD\,206610\,b}    &
  \colhead{HD\,180902\,b}    &
  \colhead{HD\,136418\,b}    &
  \colhead{HD\,212771\,b}    \\
}
\startdata
Period (d)    &
   \pB~(\peB)    &
   \pC~(\peC)    &
   \pD~(\peD)    &
   \pE~(\peE)    &
   \pH~(\peH)    &
   \pF~(\peF)    &
   \pG~(\peG)    \\
T$_p$\tablenotemark{a}~(JD)    &
   \tpB~(\tpeB)  &
   \tpC~(\tpeC)  &
   \tpD~(\tpeD)  &
   \tpE~(\tpeE)  &
   \tpH~(\tpeH)  &
   \tpF~(\tpeF)  &
   \tpG~(\tpeG)  \\
Eccentricity     & 
   \eB~(\eeB)    &
   \eC~(\eeC)    &
   \eD~(\eeD)    &
   \eE~(\eeE)    &
   \eH~(\eeH)    &
   \eF~(\eeF)    &
   \eG~(\eeG)    \\
K~(\ms)   & 
   \kB~(\keB)    &
   \kC~(\keC)    &
   \kD~(\keD)    &
   \kE~(\keE)    &
   \kH~(\keH)    &
   \kF~(\keF)    &
   \kG~(\keG)    \\
$\omega$~(deg) & 
   \omB~(\omeB)  &
   \omC~(\omeC)  &
   \omD~(\omeD)  &
   \omE~(\omeE)  &
   \omH~(\omeH)  &
   \omF~(\omeF)  &
   \omG~(\omeG)  \\
\msini~(\mjup) & 
   \msiniB~(\msinieB) &
   \msiniC~(\msinieC) &
   \msiniD~(\msinieD) &
   \msiniE~(\msinieE) &
   \msiniH~(\msinieH) &
   \msiniF~(\msinieF) &
   \msiniG~(\msinieG) \\
$a$~(AU)      &
   \arelB~(\areleB) &
   \arelC~(\areleC) &
   \arelD~(\areleD) &
   \arelE~(\areleE) &
   \arelH~(\areleH) &
   \arelF~(\areleF) &
   \arelG~(\areleG) \\
Linear trend~(\ms~yr$^{-1}$) &
   0 (fixed) &
   \trendC~(\trendeC) &
   0 (fixed) &
   0 (fixed) &
   \trendH~(\trendeH) &
   \trendF~(\trendeF) &
   0 (fixed) \\
rms~(\ms) & 
   \rmsB &
   \rmsC &
   \rmsD &
   \rmsE &
   \rmsH &
   \rmsF &
   \rmsG \\
Jitter~(\ms)  & 
5.0 &
5.0 &
5.0 &
5.0 &
5.0 &
5.0 &
5.0 \\
\chisq        & 
    \chisB &
    \chisC &
    \chisD &
    \chisE &
    \chisH &
    \chisF &
    \chisG \\
N$_{\rm obs}$ & 
    \nobsB &
    \nobsC &
    \nobsD &
    \nobsE &
    \nobsH &
    \nobsF &
    \nobsG \\
\enddata
\tablenotetext{a}{Time of periastron passage.}
\tablenotetext{b}{In cases where the eccentricity is consistent with
  $e=0$, we quote the $2\sigma$ upper limit from the MCMC analysis.}
\end{deluxetable}

\end{document}